\documentclass{article}
\pdfoutput=1 
\usepackage{arxiv}

\usepackage[utf8]{inputenc} 
\usepackage[T1]{fontenc}    
\usepackage{hyperref}       
\usepackage{url}            
\usepackage{booktabs}       
\usepackage{amsfonts}       
\usepackage{nicefrac}       
\usepackage{microtype}      
\usepackage{graphicx}

\newcommand\independent{\protect\mathpalette{\protect\independenT}{\perp}}
\def\independenT
#1#2{\mathrel{\rlap{$#1#2$}\mkern2mu{#1#2}}}

\usepackage{amsmath}
\usepackage{amsthm}

\newtheorem{corollary}{Corollary}
\newtheorem*{corollary*}{Corollary}

\newtheorem{lemma}{Lemma}
\newtheorem*{lemma*}{Lemma}

\usepackage[backend=bibtex,style=authoryear]{biblatex}
\bibliography{references}

\usepackage{tikz}

\title{Stable Feature Selection with Applications to MALDI Imaging Mass Spectrometry Data}

\author{
    \href{https://orcid.org/0000-0001-8755-584X}{\includegraphics[scale=0.06]{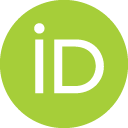}\hspace{1mm}Jonathan von Schroeder}\\
	Institute for Statistics\\
	University of Bremen\\
	Bibliothekstraße 1, D-28359 Bremen, Germany\\
	\texttt{jvs@uni-bremen.de}\\
}

\hypersetup{
pdftitle={Stable Feature Selection for MALDI Imaging Mass Spectrometry Data},
pdfsubject={math.ST},
pdfauthor={Jonathan von Schroeder},
pdfkeywords={Feature selection, U-statistic, Bootstrap, FDR control, Mass spectroscopy},
}

\begin{document}
\maketitle

\begin{abstract}
	This paper discusses an approach, based on the subsampling boostrap and FDR control, to improve the stability of feature selection. It furthermore presents the finite sample distribution of the correlation coefficient recently proposed by Chatterjee (2020) under the setting relevant for this paper. Finally an application to  matrix-assisted laser desorption/ionization (MALDI) imaging mass spectroscopy data is discussed.
\end{abstract}

\keywords{Feature selection \and U-statistic \and Bootstrap \and FDR control \and Mass spectroscopy}

\section{Introduction}
Feature selection is an important task in statistics and machine learning as it allows for dimensionality reduction and selection of features that merit further (potentially manual) analysis. This paper considers feature selection methods based on marginal test statistics. Specifically it considers the AUC (whose stability when utilized for feature selection for MALDI data was recently discussed in \cite{wilk2019}) and a newly proposed correlation coefficient and how these can be resampled. Section~\ref{sec:marginal_tests} introduces two test statistics suitable for feature selection. Section~\ref{sec:feature_selection} discusses how these univariate tests can be utilized and how the stability of the resulting feature selection procedure can be evaluated. Finally, Section~\ref{sec:application} discusses the application of the proposed methodology to matrix-assisted laser desorption/ionization (MALDI) imaging mass spectroscopy data. MALDI imaging is a mass spectrometry method suitable for high throughput imaging that shows potential for tissue typing, especially tumor (sub-)typing (c.f. \cite{boskamp2017}). This application often requires the identification of biomarkers (that is biologically meaningful features) which can be aided by stable feature selection methods.

\section{Univariate Tests for Homogeneity}\label{sec:marginal_tests}
Denote by $[n]:=\{1,\cdots,n\}$ the first $n\in\mathbb{N}$ natural numbers and let, in the following, $(X_1,Y_1),\dots,(X_n,Y_n)$ be a sample of bivariate, jointly independent random variables where (for all $i\in[n]$) $X_i\in\{0,1\}$ and $Y_i$ is real valued with absolutely continuous distribution function.

This section discusses two test statistics suitable for testing the homogeneity hypothesis (c.f. \cite[p. 51]{Dickhaus2018})
\begin{equation}
    H_0:\mathbb{P}_{Y_1}=\mathbb{P}_{Y_2}=\dots=\mathbb{P}_{Y_n}.\label{eq:homogeneity_hyp}
\end{equation}

Let $N_0:=\left|\left\{i\in[n]:X_i=0\right\}\right|$, $N_1:=n-N_0=\left|\left\{i\in[n]:X_i=1\right\}\right|$ and let $\pi=\pi(Y_1,\cdots,Y_n):[n]\rightarrow[n]$ be the (almost surely unique) random permutation such that $Y_{\pi(1)}\leq\cdots\leq Y_{\pi(n)}$. It holds that $(X_{\pi(1)},\dots,X_{\pi(n)})|N_0=n_0$ is, under $H_0$, distributed uniformly on
\begin{equation}
    B_{n_0,n_1}=\left\{x\in\{0,1\}^n\left|\sum_{i\in[n]} x_i = n_1\right.\right\}\label{eq:set_of_binary_sequences}
\end{equation} which is the set of all binary sequences of length $n$ with exactly $n_0$ zeros and $n_1:=n-n_0$ ones, since under $H_0$ $\pi$ sampled uniformly at random from the set of all permutations of $[n]$. Thus this will be the setting for which the following sections will discuss the distribution of the Mann-Whitney U statistic and that of Chatterjee's $\xi_n$.

\subsection{The Mann-Whitney U test}
The Mann-Whitney U test is a nonparametric test for the homogeneity hypothesis \eqref{eq:homogeneity_hyp}. It is based on the test statistic\begin{equation}
    U_{n_0,n_1}=U_{n_0,n_1}((X_1,Y_1),\dots,(X_n,Y_n)):=-\frac{n_0(n_0+1)}{2}+\sum_{i\in I_0}\pi^{-1}(i)
\end{equation} where $I_0:=\{i\in[n]:X_i=0\}$ and $\pi^{-1}$ denotes the inverse permutation of $\pi$ (i.e. $\pi^{-1}(i)$ is the rank of $Y_i$).

Exact and approximate $p$-values under $H_0$ can be e.g. obtained in \verb=R= using the \verb=wilcox.test= command. The expectation and variance of $U_{n_0,n_1}$ are given by $\mathbb{E}\left[U_{n_0,n_1}\right]=\frac{n_0n_1}{2}$ and $\text{Var}(U_{n_0,n_1})=\frac{n_0n_1(n+1)}{12}$ (c.f. \cite[p. 14]{lehmann1975}). Furthermore $U_{n_0,n_1}$ takes values in $[0,n_0n_1]$ (c.f. \cite[p. 9]{lehmann1975}).

In many applications $U_{n_0,n_1}$ is replaced by $\text{AUC}_{n_0,n_1}:=\frac{U_{n_0,n_1}}{n_0n_1}$ and then called the area under the ROC curve. It takes values in $[0,1]$ and has variance $\frac{n_0n_1(n+1)}{12n_0^2n_1^2}$.

\subsection{Chatterjee's $\xi$}
Recently a new coefficient of correlation was proposed in \cite{chatterjee2020}. This section discusses its properties and gives the finite sample distribution of this coefficient for the setting relevant to this paper. The coefficient is defined as\begin{align}
\xi_n(Y,X)&:=1-\frac{n\sum_{i=1}^{n-1}\left|r_{i+1}-r_i\right|}{2\sum_{i=1}^nl_i\times(n-l_i)}\label{eq:chatterjee_corr_general}
\end{align} where $r_i:=\left|\left\{j\in[n]:X_{\pi(j)}\leq X_{\pi(i)}\right\}\right|=\pi^{-1}(i)$, $l_i:=\left|\left\{j\in[n]:X_{\pi(j)}\geq X_{\pi(i)}\right\}\right|$. It is obviously symmetric, simple to compute and has the property that $\xi(Y,X):=\lim_{n\rightarrow\infty}\xi_n(Y,X)=0$ iff $Y\independent X$ and $\xi(Y,X)=1$ iff there exists a measurable $f$ such that $X=f(Y)$ almost surely (assuming $X$ is not a.s. constant, cf. \cite[Theorem 1.1.]{chatterjee2020}).

Since $X$ is a binary random variable it is immediate that\begin{align*}
r_i&=N_0+\chi\left(X_{\pi(i)}=1\right)\times N_1\\
l_i&=\chi\left(X_{\pi(i)}=0\right)\times N_0+N_1
\end{align*} and furthermore it follows that\begin{align*}
    \sum_{i=1}^Nl_i\times(n-l_i)&=\sum_{i=1}^{N_1}N_0N_1=N_0N_1^2
\end{align*} and \begin{align*}
    \sum_{i=1}^{n-1}\left|r_{i+1}-r_i\right|&=\sum_{i=1}^{n-1}N_1\chi\left(r_{i+1}\not=r_i\right)\\
    &=N_1\times\left|\left\{i\in[n-1]:X_{\pi(i+1)}\not=X_{\pi(i)}\right\}\right|.
\end{align*} Thus it follows that \eqref{eq:chatterjee_corr_general} can be written as\begin{align*}
    \xi_n(Y,X)&:=1-\frac{nN_1\times\left|\left\{i\in[n-1]:X_{\pi(i+1)}\not=X_{\pi(i)}\right\}\right|}{2N_0N_1^2}\\
    &=1-\frac{n}{2N_0N_1}\times\left|\left\{i\in[n-1]:X_{\pi(i+1)}\not=X_{\pi(i)}\right\}\right|.
\end{align*}

Since this is only well-defined if $N_0,N_1>0$ one needs to choose a value for the (uninformative) cases $N_0=0$ or $N_1=0$. Since in this case there is neither evidence against nor for independence of $X$ and $Y$ it is reasonable to choose $\xi_n(Y,X)=0$ if $N_0=0\lor N_1=0$. This choice does not affect the properties of the limit $\xi$ unless $X$ is almost surely constant.

To derive the finite sample distribution of $\xi_n$ when $X$ and $Y$ are independent and conditional on $N_0=n_0, N_1=n_1$, denote by \begin{align*}
    \tau_n(x)&=\left|\left\{i\in[n-1]\left|x_i\not=x_{i+1}\right.\right\}\right|
\end{align*} the number of 'jumps' in the binary sequence $x\in B_{n_0,n_1}$.
If $X$ is sampled uniformly at random from $B_{n_0,n_1}$, then the distribution of $\tau_n(X)$  is given by the following corollary to the results of \cite{tharrats2001}:

\begin{corollary}\label{coro:1}
    If $X$ is sampled uniformly at random from $B_{n_0,n_1}$, then the distribution of $\tau(X)$ is given by\begin{align}
       \mathbb{P}(\tau(X) = x)&=\begin{cases}
        (x+1)^2\times G_{n_0,n_1}(x+1)&\text{if $x$ is odd}\\
        (nx-x^2)\times G_{n_0,n_1}(x)&\text{if $x$ is even}
       \end{cases}\label{eq:tau_dist}
    \end{align} where $n:=n_0+n_1$, $x\in\left[2\times(n_0\land n_1)-\chi(n_0=n_1)\right]$ and where\begin{align*}
        G_{n_0,n_1}(x):=\left[2n_0n_1\times\begin{pmatrix}n\\ n_0\end{pmatrix}\right]^{-1}\times\begin{pmatrix}n_0\\ \frac{x}{2}\end{pmatrix}
        \begin{pmatrix}n_1\\ \frac{x}{2}\end{pmatrix}.
    \end{align*}
\end{corollary}
The \hyperlink{proof:coro1}{proof} is deferred to the \hyperref[a:proofs]{Appendix}.

The following lemma summarizes some properties of this distribution:

\begin{lemma}\label{lemma:1}
    The expectation of the probability distribution defined by \eqref{eq:tau_dist} is given by\begin{align*}
        \mathbb{E}[\tau(X)]&=\frac{2n_0n_1}{n}
    \end{align*}
    For $n_0=n_1=m$ it holds that $\mathbb{P}(\tau(X) = m+a)=\mathbb{P}(\tau(X) = m-a)$ for all $a\in\mathbb{N}$. Furthermore, if $n_0=n_1=m$, the variance of the probability distribution defined by \eqref{eq:tau_dist} is given by\begin{align*}
        \text{Var}(\tau(X))&=m\frac{2m^2-1}{2m-1}-m^2=\frac{m(m-1)}{2m-1}.
    \end{align*}
\end{lemma}
The \hyperlink{proof:lemma1}{proof} is deferred to the \hyperref[a:proofs]{Appendix}.

For $n_0\in[n-1]$ it holds that if $X$ and $Y$ are independent $1-\frac{n}{2n_0(n-n_0)}\tau(Z_{n_0})\equiv\xi_n(Y,X)|N_0=n_0$ since $\pi$ is then just a permutation sampled uniformly at random from the set of all permutations of $[n]$ and where $Z_{n_0}$ is sampled uniformly at random from $B_{n_0,n-n_0}$. It follows immediately that $\mathbb{E}\left[\xi_n(Y,X)|N_0=n_0\right]=0$.

\subsection{Bootstrap and U-statistics}
A popular technique to reduce the variance of an estimator of a parameter is the (subsampling) bootstrap. It is simple to demonstrate that the ordinary bootstrap (i.e. resampling with replacement instead of subsampling) is inappropriate for $\xi_n$: If $(X_1^*,Y_1^*),\cdots,(X_n^*,Y_n^*)$ is a bootstrap sample (sampled with replacement), then\begin{align*}
    \mathbb{E}_{\mathbb{P}^*}\left[\left|\left\{i\in[n-1]:X_{\pi(i+1)}^*\not=X_{\pi(i)}^*\right\}\right|\right]&=n-1-\mathbb{E}\left[\left|\left\{i\in[n-1]:X_{\pi(i+1)}^*= X^*_{\pi(i)}\right\}\right|\right]\\
    &\leq \left(n-1\right)\times\left(1-\left(1-n^{-1}\right)^n\right)
\end{align*} since the expected fraction of non-unique points in the bootstrap sample is $\left(1-n^{-1}\right)^n$. Consequently
\begin{align*}
    \mathbb{E}_{\mathbb{P}^*}[\xi_n(X^*,Y^*)]\geq1-\frac{n}{2n_1n_2}\times(n-1)\times\left(1-\left(1-n^{-1}\right)^n\right)
\end{align*} and thus for $O\left(n^{-1}n_1\right)=O\left(n^{-1}n_2\right)=O(1)$ it follows that $\lim_{n\rightarrow\infty} \mathbb{E}_{\mathbb{P}^*}[\xi_n(X^*,Y^*)]\geq e^{-1}$. But it follows from \cite[Theorem 1.1]{chatterjee2020} that $\lim_{n\rightarrow\infty} \mathbb{E}[\xi_n(X,Y)]=0$ if $X$ and $Y$ are independent and therefore the bootstrap fails to be asymptotically unbiased in this case. Sampling without replacement does not cause artificial ties and therefore does not suffer from this issue.

If all subsamples of a fixed size $m$ are evaluated and the resampled statistic is symmetric, then the average over all such subsamples is an U-statistic which is a type of unbiased and asymptotically normal test statistic that was introduced by \cite{hoeffding1948}. Since both of the previously discussed test statistics are symmetric functions of the sample one can use them as the kernel of a U statistic (c.f. \cite[Definition 1.1]{bose2018}) and define\begin{align*}
    T^{\text{AUC}}_{(m,n)}&:=\binom{n_0+n_1}{m}^{-1}\sum_{1\leq i_1<\dots<i_m\leq n}\text{AUC}_{\tilde n_0,\tilde n_1}\left((X_{i_1},Y_{i_1}),\dots,(X_{i_m},Y_{i_m})\right)\\
    T^{\xi}_{(m,n)}&:=\binom{n_0+n_1}{m}^{-1}\sum_{1\leq i_1<\dots<i_m\leq n}\xi_{\tilde n_0,\tilde n_1}\left((X_{i_1},Y_{i_1}),\dots,(X_{i_m},Y_{i_m})\right)
\end{align*} where $\tilde n_1=\tilde n_1(i_1,\cdots,i_m):=\sum_{j=1}^m X_{i_j}$ and $\tilde n_0=\tilde n_0(i_1,\cdots,i_m):=m-\tilde n_1$. In practice $\binom{n_0+n_1}{m}^{-1}$ will be so large for many applications, that an exact evaluation of $T^{\text{AUC}}_{(m,n)}$ and $T^{\xi}_{(m,n)}$ is infeasible. It is however possible to replace these test statistics by Monte Carlo approximations. To this end denote by $I^{(1)},\cdots,I^{(\ell)}$ a sample of size $\ell$ drawn uniformly at random from $\left\{(i_1,\cdots,i_m)\in[n]^m:1\leq i_1<\dots<i_m\leq n\right\}$. Then\begin{align}
    \tilde T^{\text{AUC},\ell}_{(m,n)}&:=\ell^{-1}\sum_{j=1}^\ell\text{AUC}_{\tilde n_0,\tilde n_1}\left(\left(X_{I_1}^{(j)},Y_{I_1}^{(j)}\right),\dots,\left(X_{I_m}^{(j)},Y_{I_m}^{(j)}\right)\right)\label{eq:approx_auc_sub}\\
    \tilde T^{\xi,\ell}_{(m,n)}&:=\ell^{-1}\sum_{j=1}^\ell\xi_{\tilde n_0,\tilde n_1}\left(\left(X_{I_1}^{(j)},Y_{I_1}^{(j)}\right),\dots,\left(X_{I_m}^{(j)},Y_{I_m}^{(j)}\right)\right)\label{eq:approx_xi_sub}
\end{align} are randomized approximations of the previously defined quantities that converge almost surely as $\ell\rightarrow\infty$ due to the strong law of large numbers. $P$-values for these test statistics can be approximated, according to the methodology described in \cite{phipson2010} (which is implemented in the \verb=R=-package \verb=statmod=).

\section{Feature Selection}\label{sec:feature_selection}
When performing feature selection based on some ranking of the features (e.g. in terms of the observed AUC) it is challenging to decide how many features to select: Selecting only very few features might drop very important features (e.g. degrading classification performance) whereas selecting too many might not reduce the dimensionality of the problem sufficiently and/or keep many irrelevant features. If marginal $p$-values can be obtained (which is the case for the AUC and $\xi_n$ as well as their resampled counterparts) this trade-off can be tackled by using a procedure to control the false discovery rate (FDR) when performing multiple comparisons. This section gives a short introduction to FDR control and discusses how to evaluate the stability of the proposed feature selection method, which, given a sample $Y\in\mathbb{R}^{n\times p}$ and $X\in\{0,1\}^{n}$, consists of two steps:
\begin{enumerate}
    \item Calculate marginal test statistics and $p$-values, that is the j-th $p$-value (for $j=1,\cdots,p$) is calculated from the sample $(Y_{\cdot,j},X)$ consisting of the j-th column of $Y$ and the random vector $X$.
    \item Select features (that is $I\subset [p]$) according to the Benjamin-Yekutieli procedure, controlling the FDR at the desired level $\alpha$
\end{enumerate}

When calculating the test statistics and $p$-values in step 1 it is necessary to decide how $I^{(1)},\cdots,I^{(l)}$ are drawn. If, for each a marginal, an independent sample is drawn the procedure is called independent-component bootstrap. It has been argued by \cite{hall2009} that this can be reasonable even if the component vectors are not independent and this approach was successfully applied by \cite{neumann2021} to estimate the proportion of true null hypotheses under dependency. Nonetheless using the same sample for all marginals seems more appropriate for the goal of this paper:
\begin{itemize}
    \item It turns out that, when calculating \eqref{eq:approx_auc_sub} or \eqref{eq:approx_xi_sub} for large data sets, much of the computational effort is spent on generating pseudo-random numbers. Thus it is desirable to use as few pseudo-random numbers as possible.
    \item From a theoretical point of view using different samples could lead to different values of the test statistic and the estimated $p-$values for very similar marginal samples. This is obviously undesirable and therefore it is preferable to use the same approximation $\tilde T^{\text{AUC},\ell}_{(m,n)}$ (or $\tilde T^{\xi,\ell}_{(m,n)}$) for all of the simultaneous tests.
\end{itemize}

Furthermore it is necessary choose an appropriate $m$. The heuristic proposed in \cite{bickel2008} suggested, for the examples discussed in the following section, that choosing $m$ very small is inappropriate, which is unsurprising since the number of values the resampled test statistic can  take is very small if $m$ is very small. Since, by the results of the previous section, the variance of the AUC and $\xi$ goes to zero if $n_0,n_1\rightarrow\infty$ it might as first seem like it would be a good idea to choose an $m$ on the same order of magnitude as the sample size $n$. However, as $\ell\rightarrow\infty$ the variance of e.g. $\tilde T^{\xi,\ell}_{(m,n)}$ depends only on the covariance between $Z_1:=\xi_{\tilde n_0,\tilde n_1}\left(\left(X_{I_1}^{(1)},Y_{I_1}^{(1)}\right),\dots,\left(X_{I_m}^{(1)},Y_{I_m}^{(1)}\right)\right)$ and $Z_2:=\xi_{\tilde n_0,\tilde n_1}\left(\left(X_{I_1}^{(1)},Y_{I_1}^{(2)}\right),\dots,\left(X_{I_m}^{(1)},Y_{I_m}^{(2)}\right)\right)$ since\begin{align*}
    \text{Var}\left(\tilde T^{\xi,\ell}_{(m,n)}\right)&=\ell^{-1}\text{Var}(V_1)+\frac{\ell-1}{\ell}\text{Cov}(V_1,V_2).
\end{align*} This covariance is, however, increasing in $m$ (as has been checked numerically for the examples in the next section) and therefore a small, but not extremely small choice of $m$ seems to be best.

To allow for a parallelized implementation of the proposed approach that yields, for a given seed (and up to numerical accuracy), reproducible results, the random number generator proposed \cite{LEcuyer1999} and implemented in \verb=C++= as described in \cite{LEcuyer2002} is utilized.

When simultaneously testing multiple hypotheses a type-I error can be made for each hypothesis. Therefore it is, in general, not sufficient to control the type-I error at the nominal significance level for each of the tests. Instead $V$, the (random) number of incorrectly rejected null hypotheses, needs to be controlled. Controlling the probability $\mathbb{P}(V\geq1)$ is called family-wise error rate (FWER) control. For large scale problems, methods controlling the FWER are usually too strict. A good alternative is controlling the FDR, which is the expected value of the ratio $V/R$ (where the random variable $R$ is the total number of rejected null hypotheses). In \cite{benjamini2001} a procedure, that is usually called the Benjamini-Yekutieli (BY) procedure, is proposed which controls the FDR under arbitrary dependencies between the hypotheses under consideration. Since the dependency structure is unknown for the application considered in this paper, this is the procedure that will be utilized. Of course the proposed approach works with other methods for FDR control as well, when these are appropriate.

Inspired by the work of \cite{meinshausen2010} the stability of the proposed feature selection method will be evaluated in terms of the (estimated) number of features that have a very hight selection probability (cf. \cite[Definition 1]{meinshausen2010}). This is done by considering a partition of the sample into $k$ disjoint subsamples $[n]=\bigcup_{i=1}^k B_i$ (as would be done for $k$-fold cross validation) and the number \begin{equation*}
    S(M_s):=\left|\bigcap_{i=1}^k M_s(B_i)\right|
\end{equation*} where $M_s(B_i)$ are the indices of the $s$ top ranked features (based on either on the AUC or $\xi$), i.e. $\left|M_s(B_i)\right|=s$. That is $S(M_s)\in[s]$ counts the number of features thateawre selected in each of the cross validation folds.

\section{Application to MALDI Imaging Mass Spectrometry Data}\label{sec:application}
This section demonstrates the proposed methodology by reanalyzing the data-set presented in \cite{kriegsmann2016} for which 5 biomarkers (that is biologically meaningful m/z values) have been identified (cf. \cite[Supplementary Table 1]{kriegsmann2016}). This data-set was subsequently analyzed by \cite{boskamp2017}, \cite{behrmann2017} and \cite{leuschner2018}. For a detailed description of the data set as well the pre-processing see either of these papers. To make a comparison between the results presented in this paper and those of \cite{behrmann2017} and \cite{leuschner2018} simple the same 4-fold (respectively 8-fold) cross-validation strategy was employed. The pre-processing was performed as described in these papers with the following changes:
\begin{itemize}
    \item In both cases a novel multiplicative-trend normalization (which was performed on a per core basis and using a running median filter with fixed window size $301$) was performed before the usual total ion count (TIC) normalization.
    \item For the task considered in \cite{behrmann2017} the spectra were additionally resampled to intervals of 1 Da width resulting in 3046 m/z channels. This was done to achieve a number of features similar to that on which the analysis in \cite{leuschner2018} was based.
\end{itemize}

\begin{figure}[!htb]
	\centering
    \resizebox{.95\linewidth}{!}{\input{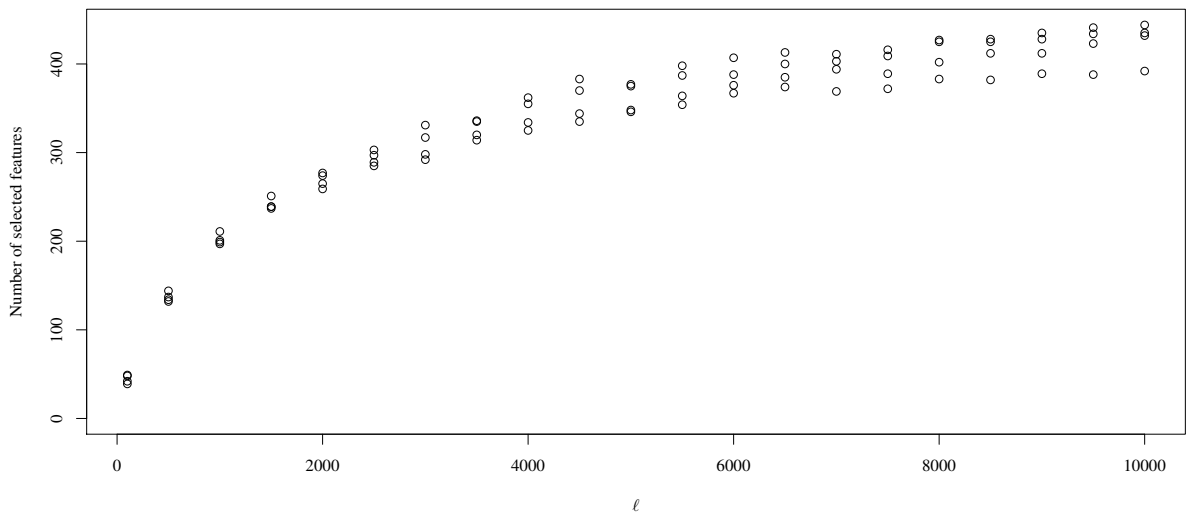}}
	\caption{The number of features selected by the resamples $\xi$ when controlling the FDR at level $\alpha=0.15$ (using the BY procedure) as function of the number of subsamples $\ell$ and for the cross validation folds of the ADSQ (Spot) task of \cite{behrmann2017}. Resampling was performed with subsamples of size $m=50$.}
	\label{fig:rejections_fdr}
\end{figure}

Table~\ref{tab:bacc} demonstrates that the proposed method is, when combined with a \verb=randomForest= classifier (cf. \cite{randomForest}) with standard settings, competitive with the state of the art for the reanalyzed data-set. For the first task $\ell$ was chosen very small on purpose because otherwise (as was the case for the non-resampled $\xi$) all of the features would have been chosen (when controlling the FDR at $\alpha=0.15$). This suggests that the hyperparameter $\ell$ affects the ability of the test to reject the null-hypothesis. Indeed this effect can be seen in Figure~\ref{fig:rejections_fdr}.

\begin{figure}[!htb]
	\centering
    \resizebox{.95\linewidth}{!}{\input{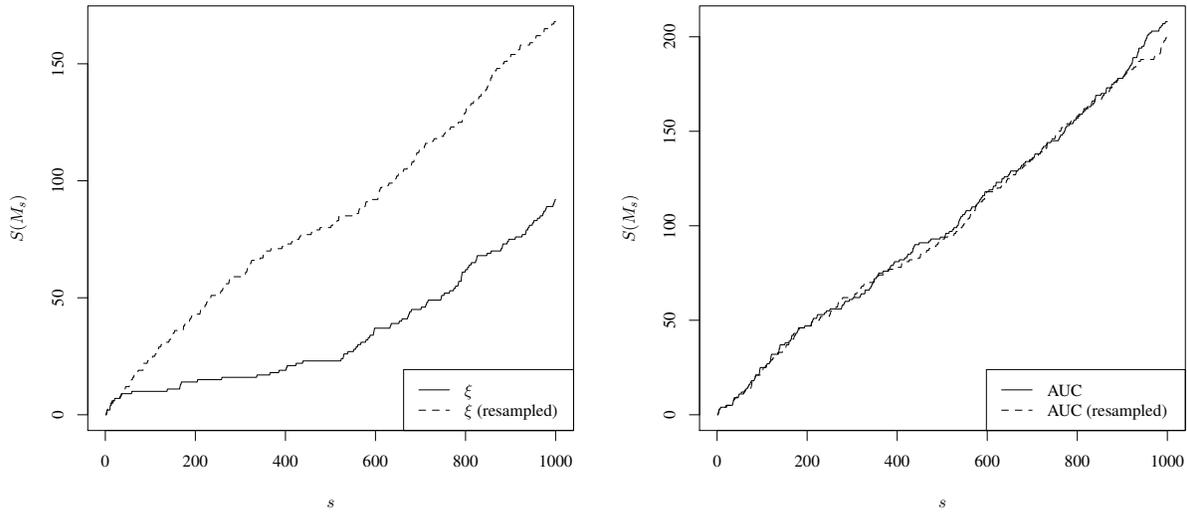}}
	\caption{The number of stably selected features (as a function of the number of selected features $s$) for the ADSQ (Spot) task of \cite{behrmann2017}. Resampling was performed with subsample size $m=50$ and $\ell=1000$ subsamples.}
	\label{fig:stability}
\end{figure}

In Figure~\ref{fig:stability} it can be seen that the resampling improves the selection stability of $\xi$ while leaving that of the AUC mostly unchanged.

\begin{table}
    \caption{\label{tab:bacc}Average balanced accuracy achieved over the 8 (Task \cite{leuschner2018}) /  4 (Task ADSQ (Spot)) cross validation folds. The method IsotopeNet is that of \cite{behrmann2017} and Flog\_int is one of the methods described in \cite{leuschner2018}. The other methods are those described in this paper. Resampling was performed with subsample size $m=50$ and $\ell=100$ ($\ell=1000$) for the first (second) task. All $p$-values used for the FDR control in the feature selection procedures were estimated using $10^5$ Monte Carlo samples.}
	\centering
	\begin{tabular}{lcc}
		\toprule
		&\multicolumn{2}{c}{Task}                   \\
		\cmidrule(r){2-3}
		Method     & \cite{leuschner2018} & ADSQ (Spot) \cite{behrmann2017} \\
		\midrule
		IsotopeNet &   & 0.845 \\
		Flog\_int & 0.904 & \\
		AUC and Random Forest  & 0.927 & 0.870 \\
        AUC (resampled) and Random Forest  &  0.927 &  0.867  \\
        $\xi$ and Random Forest  & 0.926 & 0.848 \\
        $\xi$ (resampled) and Random Forest  & 0.930 & 0.866 \\
		\bottomrule
    \end{tabular}
\end{table}

\section{Discussion}

This paper introduced an approach to feature selection that is based on the subsampling bootstrap and FDR control. It furthermore presented the finite sample distribution of the correlation coefficient recently proposed by \cite{chatterjee2020} under the setting relevant for this paper. Finally an application to MALDI mass spectroscopy data was presented. The results of this application suggest that the number of subsamples $\ell$ plays the role of a hyper-parameter that controls the ability of the test to reject the null-hypothesis. It would therefore be interesting to, in future research, investigate how this is related to the effect size for specific alternatives and if there is a principled way to calibrate $\ell$ such that the test becomes less sensitive towards effect sizes smaller than some minimal effect.

Furthermore it could be interesting to consider methods to control the resampling risk introduced by Monte Carlo approximation of the $p$-values, that is the risk that for some hypothesis the test decision is different from the one based on the theoretical $p$-value. Towards this end \cite{gandy2009} proposed a sequential implementation of Monte Carlo tests with uniformly bounded resampling risk, which could be applied to estimate the $p-$values required by the method proposed in this paper. However, convergence has been observed to be extremely slow and therefore the application is unfortunately not straightforward.

\section*{Acknowledgments}
Jonathan von Schroeder is supported by the Deutsche Forschungsgemeinschaft (DFG) within the framework of RTG 2224
"$\pi^3$: Parameter Identification - Analysis, Algorithms, Applications".

\nocite{r_manual}
\printbibliography

@misc{tharrats2001,
    author = {J. Tharrats},
    title = {Structure of Binary Sequences},
    year = {2001},
    eprint = {arXiv:math/0109182},
}

@article{chatterjee2020,
  doi = {10.1080/01621459.2020.1758115},
  year = {2020},
  month = may,
  publisher = {Informa {UK} Limited},
  pages = {1--21},
  author = {Sourav Chatterjee},
  title = {A New Coefficient of Correlation},
  journal = {Journal of the American Statistical Association}
}

@article{hoeffding1948,
    author = "Hoeffding, Wassily",
    doi = "10.1214/aoms/1177730196",
    fjournal = "Annals of Mathematical Statistics",
    journal = "Ann. Math. Statist.",
    month = "09",
    number = "3",
    pages = "293--325",
    publisher = "The Institute of Mathematical Statistics",
    title = "A Class of Statistics with Asymptotically Normal Distribution",
    volume = "19",
    year = "1948"
}

@article{benjamini2001,
    author = "Benjamini, Yoav and Yekutieli, Daniel",
    doi = "10.1214/aos/1013699998",
    fjournal = "Annals of Statistics",
    journal = "Ann. Statist.",
    number = "4",
    pages = "1165--1188",
    publisher = "The Institute of Mathematical Statistics",
    title = "The Control of the False Discovery Rate in Multiple Testing Under Dependency",
    volume = "29",
    year = "2001"
}

@article{meinshausen2010,
  doi = {10.1111/j.1467-9868.2010.00740.x},
  year = {2010},
  publisher = {Wiley},
  volume = {72},
  number = {4},
  pages = {417--473},
  author = {Nicolai Meinshausen and Peter B\"{u}hlmann},
  title = {Stability Selection},
  journal = {Journal of the Royal Statistical Society: Series B (Statistical Methodology)}
}

@book{Dickhaus2018,
  doi = {10.1007/978-3-319-76315-6},
  year = {2018},
  publisher = {Springer International Publishing},
  author = {Thorsten Dickhaus},
  title = {Theory of Nonparametric Tests}
}

@book{bose2018,
  doi = {10.1007/978-981-13-2248-8},
  year = {2018},
  publisher = {Springer Singapore},
  author = {Arup Bose and Snigdhansu Chatterjee},
  title = {U-Statistics, Mm-Estimators and Resampling}
}

@article{hall2009,
  doi = {10.1214/09-aos699},
  year = {2009},
  month = dec,
  publisher = {Institute of Mathematical Statistics},
  volume = {37},
  number = {6B},
  pages = {3929--3959},
  author = {Peter Hall and Hugh Miller},
  title = {Using the Bootstrap to Quantify the Authority of an Empirical Ranking},
  journal = {The Annals of Statistics}
}

@article{neumann2021,
  doi = {10.1016/j.jspi.2020.04.011},
  year = {2021},
  month = jan,
  publisher = {Elsevier {BV}},
  volume = {210},
  pages = {76--86},
  author = {Andr{\'{e}} Neumann and Taras Bodnar and Thorsten Dickhaus},
  title = {Estimating the Proportion of True Null Hypotheses Under Dependency: A Marginal Bootstrap Approach},
  journal = {Journal of Statistical Planning and Inference}
}

@article{phipson2010,
  doi = {10.2202/1544-6115.1585},
  year = {2010},
  publisher = {Walter de Gruyter},
  volume = {9},
  number = {1},
  author = {Belinda Phipson and Gordon K Smyth},
  title = {Permutation P-values Should Never Be Zero: Calculating Exact P-values When Permutations Are Randomly Drawn},
  journal = {Statistical Applications in Genetics and Molecular Biology}
}

@article{bickel2008,
 URL = {http://www.jstor.org/stable/24308525},
 author = {Peter J. Bickel and Anat Sakov},
 journal = {Statistica Sinica},
 number = {3},
 pages = {967--985},
 publisher = {Institute of Statistical Science, Academia Sinica},
 title = {On the Choice of m in the m out of n Bootsrap and Confidencen Bounds for Extrema},
 volume = {18},
 year = {2008},
 month = jul
}

@article{leuschner2018,
  doi = {10.1093/bioinformatics/bty909},
  year = {2018},
  month = nov,
  publisher = {Oxford University Press ({OUP})},
  volume = {35},
  number = {11},
  pages = {1940--1947},
  author = {Johannes Leuschner and Maximilian Schmidt and Pascal Fernsel and Delf Lachmund and Tobias Boskamp and Peter Maass},
  title = {Supervised non-negative matrix factorization methods for {MALDI} imaging applications},
  journal = {Bioinformatics}
}

@article{behrmann2017,
  doi = {10.1093/bioinformatics/btx724},
  year = {2018},
  publisher = {Oxford University Press ({OUP})},
  volume = {34},
  number = {7},
  pages = {1215--1223},
  author = {Jens Behrmann and Christian Etmann and Tobias Boskamp and Rita Casadonte and J\"{o}rg Kriegsmann and Peter Maa$\upbeta$},
  title = {Deep Learning For Tumor Classification in Imaging Mass Spectrometry},
  journal = {Bioinformatics}
}

@article{kriegsmann2016,
  doi = {10.1074/mcp.m115.057513},
  year = {2016},
  month = jul,
  publisher = {American Society for Biochemistry {\&} Molecular Biology},
  volume = {15},
  number = {10},
  pages = {3081--3089},
  author = {Mark Kriegsmann and Rita Casadonte and J\"{o}rg Kriegsmann and Hendrik Dienemann and Peter Schirmacher and Jan Hendrik Kobarg and Kristina Schwamborn and Albrecht Stenzinger and Arne Warth and Wilko Weichert},
  title = {Reliable Entity Subtyping in Non-small Cell Lung Cancer by Matrix-assisted Laser Desorption/Ionization Imaging Mass Spectrometry on Formalin-fixed Paraffin-embedded Tissue Specimens},
  journal = {Molecular {\&} Cellular Proteomics}
}

@article{boskamp2017,
  doi = {10.1016/j.bbapap.2016.11.003},
  year = {2017},
  month = jul,
  publisher = {Elsevier {BV}},
  volume = {1865},
  number = {7},
  pages = {916--926},
  author = {Tobias Boskamp and Delf Lachmund and Janina Oetjen and Yovany Cordero Hernandez and Dennis Trede and Peter Maass and Rita Casadonte and J\"{o}rg Kriegsmann and Arne Warth and Hendrik Dienemann and Wilko Weichert and Mark Kriegsmann},
  title = {A New Classification Method for {MALDI} Imaging Mass Spectrometry Data Acquired on Formalin-Fixed Paraffin-Embedded Tissue Samples},
  journal = {Biochimica et Biophysica Acta ({BBA}) - Proteins and Proteomics}
}

@article{LEcuyer1999,
  doi = {10.1287/opre.47.1.159},
  year = {1999},
  month = feb,
  publisher = {Institute for Operations Research and the Management Sciences ({INFORMS})},
  volume = {47},
  number = {1},
  pages = {159--164},
  author = {Pierre L'Ecuyer},
  title = {Good Parameters and Implementations for Combined Multiple Recursive Random Number Generators},
  journal = {Operations Research}
}

@article{LEcuyer2002,
  doi = {10.1287/opre.50.6.1073.358},
  year = {2002},
  month = dec,
  publisher = {Institute for Operations Research and the Management Sciences ({INFORMS})},
  volume = {50},
  number = {6},
  pages = {1073--1075},
  author = {Pierre L'Ecuyer and Richard Simard and E. Jack Chen and W. David Kelton},
  title = {An Object-Oriented Random-Number Package with Many Long Streams and Substreams},
  journal = {Operations Research}
}

@incollection{wilk2019,
  doi = {10.1007/978-3-030-29885-2_8},
  year = {2019},
  month = aug,
  publisher = {Springer International Publishing},
  pages = {82--93},
  author = {Agata Wilk and Marta Gawin and Katarzyna Fr{\k{a}}tczak and Piotr Wid{\l}ak and Krzysztof Fujarewicz},
  title = {On Stability of Feature Selection Based on {MALDI} Mass Spectrometry Imaging Data and Simulated Biopsy},
  booktitle = {Advances in Intelligent Systems and Computing}
}

@article{randomForest,
    title = {Classification and Regression by randomForest},
    author = {Andy Liaw and Matthew Wiener},
    journal = {R News},
    year = {2002},
    volume = {2},
    number = {3},
    pages = {18-22},
    url = {https://CRAN.R-project.org/doc/Rnews/},
}

@Manual{r_manual,
    title = {R: A Language and Environment for Statistical Computing},
    author = {{R Core Team}},
    organization = {R Foundation for Statistical Computing},
    address = {Vienna, Austria},
    year = {2020},
    url = {https://www.R-project.org/},
}

@article{gandy2009,
  doi = {10.1198/jasa.2009.tm08368},
  year = {2009},
  publisher = {Informa {UK} Limited},
  volume = {104},
  number = {488},
  pages = {1504--1511},
  author = {Axel Gandy},
  title = {Sequential Implementation of Monte Carlo Tests With Uniformly Bounded Resampling Risk},
  journal = {Journal of the American Statistical Association}
}

@book{lehmann1975,
  title={Nonparametrics: Statistical Methods Based on Ranks},
  author={Erich Leo Lehmann and D'Abrera, H.J.M.},
  year={1975},
  publisher={Holden-Day}
}

\appendix
\section{Proofs}\label{a:proofs}

\begin{proof}[\hypertarget{proof:coro1}{Proof of Corollary~\ref{coro:1}}]
    Simply note that $\tau$ in \cite{tharrats2001} and $\tau$ in this publication equal if $x_1=x_n$ and the $\tau$ in this publication is one smaller otherwise. Thus one only needs to distinguish these. This is fortunately simple since the latter kind of sequences correspond to the loops (where on also needs to consider those loops that start with a block of ones). There are (in the notation of Tharrats (2001)) exactly $2{m-1\choose h-1}{n-1\choose h-1}$ such loops which yields the desired formula. Thus it follows that\begin{align*}
        \mathbb{P}(\tau(X) = x)&=\begin{pmatrix}n\\ n_1\end{pmatrix}^{-1}\begin{cases}
        2\times
        \begin{pmatrix}n_1-1\\ \frac{x+1}{2}-1\end{pmatrix}
        \begin{pmatrix}n_2-1\\ \frac{x+1}{2}-1\end{pmatrix}&\text{if $x$ is odd}\\
        \left(\dfrac{2n}{x}-2\right)\times
        \begin{pmatrix}n_1-1\\ \frac{x}{2}-1\end{pmatrix}
        \begin{pmatrix}n_2-1\\ \frac{x}{2}-1\end{pmatrix}&\text{if $x$ is even}
       \end{cases}\\
       &=\begin{pmatrix}n\\ n_1\end{pmatrix}^{-1}\begin{cases}
        \dfrac{(x+1)^2}{2n_1n_2}\times
        \begin{pmatrix}n_1\\ \frac{x+1}{2}\end{pmatrix}
        \begin{pmatrix}n_2\\ \frac{x+1}{2}\end{pmatrix}&\text{if $x$ is odd}\\
        \dfrac{nx-x^2}{2n_1n_2}\times
        \begin{pmatrix}n_1\\ \frac{x}{2}\end{pmatrix}
        \begin{pmatrix}n_2\\ \frac{x}{2}\end{pmatrix}&\text{if $x$ is even}
       \end{cases}
    \end{align*} which yields the desired result.
\end{proof}

\begin{proof}[\hypertarget{proof:lemma1}{Proof of Lemma~\ref{lemma:1}}]
    To show the symmetry write \begin{align*}
        x=m\pm a=\begin{cases}
        m+a&\text{if }x\geq a\\
        m-a&\text{if }x<0
    \end{cases}
    \end{align*} for $x\in[2m-1]$ and $a>0$. Then \begin{align*}
        \mathbb{P}(\tilde\tau_{n,n_1,n_2} = x)&=\begin{cases}
            (m\pm a+1)^2\times G_{m,m}(m\pm a+1)&\text{if $x$ is odd}\\
            (m^2-a^2)\times G_{m,m}(m\pm a)&\text{if $x$ is even}
           \end{cases}
    \end{align*}

    Thus, for $m\in\mathbb{N},a\in\mathbb{N}$ and $m\pm a$ odd it suffices to show that $(m+a+1)^2\times G_{m,m}(m+a+1)=(m-a+1)^2\times G_{m,m}(m-a+1)$.
    To verify this claim one only needs to check the identity\begin{align*}
        (m+a+1)\times\begin{pmatrix}m\\ \frac{m+a+1}{2}\end{pmatrix}=(m-a+1)\times\begin{pmatrix}m\\ \frac{m-a+1}{2}\end{pmatrix}.
    \end{align*} To this end it is helpful to distinguish two cases:\begin{itemize}
        \item If $m$ is odd (and thus $a$ is even), then it is sufficient to verify that\begin{align*}
            (m+a+1)\times\begin{pmatrix}2m+1\\ m+a+1\end{pmatrix}=(m-a+1)\times\begin{pmatrix}2m+1\\ m-a+1\end{pmatrix}
        \end{align*} for arbitrary $m,a\in\mathbb{N}$.
         \item If $m$ is even (and thus $a$ is odd), then it is sufficient to verify that\begin{align*}
            (m+a+1)\times\begin{pmatrix}2m\\ m+a+1\end{pmatrix}=(m-a)\times\begin{pmatrix}2m\\ m-a\end{pmatrix}
        \end{align*} for arbitrary $m,a\in\mathbb{N}$.
    \end{itemize}

    Furthermore for $m\pm a$ even it holds, that $G_{m,m}(m+a)=G_{m,m}(m-a)$ and thus the density of $\tau(X)$ is symmetric about $m$ and therefore $\mathbb{E}[\tau(X)]=m$.

    $\mathbb{E}[\tau(X)]=\frac{2n_0n_1}{n}$ holds for $n_0\not=n_1$ since it holds that\begin{align*}
        \mathbb{E}[\tau(X)]&=\sum_{x=1}^{2(n_0\land n_1)}x\mathbb{P}(\tau(X)=x)\\
        &=\sum_{x=1}^{n_0\land n_1} 2x\mathbb{P}(\tau(X)=2x)+(2x-1)\mathbb{P}(\tau(X)=2x-1)\\
        &=\sum_{x=1}^{n_0\land n_1} G_{n_0,n_1}(2x)\times\left[2x\left(2nx-(2x)^2\right)+(2x-1)\cdot(2x-1+1)^2\right]\\
        &=4(n-1)\times\sum_{x=1}^{n_0\land n_1}x^2G_{n_0,n_1}(2x)=\frac{(n-1)}{n_0n_1\binom{n}{n_0}}\times\sum_{x=1}^{n_0\land n_1}x^2\binom{n_0}{x}\binom{n_1}{x}\\
        &=\frac{2n_0n_1}{n}
    \end{align*} where the last equality follows from the relation \begin{align*}
        \sum_{h=1}^{m\land n}h^2\binom{m}{h}\binom{n}{h}&=\frac{m^2n^2}{(m+n)(m+n-1)}\binom{m+n}{m}
    \end{align*} proved in Tharrats (2001) on p. 8.

    For $n_0=n_1=m$ it holds that\begin{align*}
        \mathbb{E}[\tau(X)^2]&=\sum_{x=1}^{2m-1}x^2\mathbb{P}(\tau(X)=x)\\
        &=\sum_{x=1}^{m}G_{m,m}(2x)\times\left[(2x)^2\times\left(2x\cdot (2m)-(2x)^2\right)+(2x-1)^2\times(2x)^2\right]\\
        &=\frac{2}{m^2\binom{2m}{m}}\times\left[\sum_{x=1}^m\left(x^2+4(m-1)x^3\right)\binom{m}{x}^2\right]\\
        &=\frac{2}{m^2}\times\left[\frac{m^3}{2(2m-1)}+4(m-1)\frac{m^3}{4(2m-1)}(1+m)\right]\\
        &=\frac{m}{2m-1}+\frac{2(m-1)m(m+1)}{2m-1}=\frac{m(2m^2-1)}{2m-1}
    \end{align*} where the exact relations (13,2) and (13,3) of  Tharrats (2001) where used. Thus the formula for the variance follows from $\text{Var}(\tau(X))=\mathbb{E}[\tau(X)^2]-m^2$.
\end{proof}

\end{document}